\documentclass[preprint]{revtex4}
\usepackage{graphicx}

\begin{document}

\vspace{1cm}

 \centerline {\Large \bf  squeezed number eigenstate of XYZ
 Heisenberg}
 \centerline {\Large \bf antiferromagnetics under an magnetic field}
 \vspace{1cm}
\centerline{Bing-Hao Xie$^1$, Shuo Jin$^2$ and Wei-Xian Yan$^3$}

 \vspace{0.5cm} {\normalsize
\centerline{ 1. Beijing Information Technology Institute, Beijing
100101, China}

{\normalsize \centerline {2.  Department of Physics, Beihang
University, Beijing 100083, China}

{\normalsize \centerline {3.  Institute of Theoretical Physics,
Shanxi University, Taiyuan, 030006, China}

\vspace{1.5cm}

\centerline{\bf Abstract}
 By using an
algebraic diagonalization method, the XYZ
 Heisenberg  antiferromagnetics under an external magnetic
field is studied in the framework of spin-wave theory. The energy
eigenstates are shown to be squeezed number states and the energy
eigenvalues are obtained in some cases. Some quantum properties of
the energy eigenstates, and the connection of the model with the
two-mode coupled harmonic oscillators are also discussed.}
 \vspace{0.2cm}


 \pagebreak

 \section {Introduction}
 The Heisenberg model [Heisenberg 1928; Anderson 1951], describing spin lattices with short-range interactions,
 plays an indispensable role in the study of the magnetism of solids, and many other related physical fields,
 such as  non-local spin systems, quantum dot, nuclear spin, and so on [Loss 1988; Burkard 1999; Imamoglu 1999].
  According to the sign of interaction intensity $J$,
 this model can be classified as ferromagnetic ($J>0$) type  and
antiferromagnetic ($J<0$) type, the latter is much complicated
than the former in the eigenstate structure. Based on the
interaction intensity along different space directions, the
Heisenberg model can be classified as XXX, XXZ, XYZ, where the
former two cases  can be regarded as the special case of the
latter one. In theoretical study, a basic task is to do the
digonalization of the Hamiltonian. However, due to the complicacy
of the many body problem, getting the exact solution is very
difficult, and till now only some special cases can be exactly
solved, such as one dimension XXX antiferromagnetic chain using
the Bethe-Ansatz method [Bethe 1930]. In practice, the spin-wave
theory is widely employed as an acceptable approximation, and the
"N\'eel states" is regarded to be the approximate ground states of
the antiferromagnetic case. For XXZ antiferromagnetic Heisenberg
model, spin-wave theory has been well studied in an analytical way
which can be found in standard textbooks [Callaway 1976]. However,
for XYZ case, as far as we know, there is no result parallel to
XXZ case in literature. The reason lies in that, although it's not
very difficult in principle to deal with the problem through the
traditional Bogoliubov-valatin transformation [Bogoliubov 1958;
Valatin 1958], the solution form is too complicated, to be
intuitive in physics, and one have to rely on pure numerical
analysis, leading to the lack of physical transparency.

As is well known, algebraic method is very important in studying
the system with certain dynamical group [Kim 1988; Artoni 1991;
Pan 2001, 2004]. In this paper, an algebraic diagonalization
method will be employed to do the digonalization of XYZ
antiferromagnetics under an external magnetic field in spin-wave
framework. The energy eigenstates which turn out to be squeezed
number states [Yuen 1976; Kim 1989; Nieto 1997] and the
corresponding eigenvalues in some cases will be obtained. The
paper is organized as follows. Firstly, the algebraic
diagonalization method is briefly introduced in section II. In
section III, we study the system and give the result of
eigenstates and eigenvalues in some cases. The relation between
algebraic method and Bogoliubov method is also discussed. In
section IV, we discuss some statistical properties of the
eigenstates. In section V, we investigate the relation of the
system and the two-mode coupled harmonic oscillators. Finally,
some concluding remarks are given.

\section { An algebraic
diagonalization method}
 The algebraic diagonalization
method introduced here is applicable for a class of rather general
systems whose Hamiltonian possesses dynamical semisimple
Lie-algebra structure, i.e.
   \begin{equation}
   \label{e1}
    H=\sum_i\varepsilon_i
    H_i+\sum_{\alpha}(\lambda_{\alpha}E_{\alpha}+\lambda_{\alpha}^*E_{-\alpha}),
   \end{equation}
   where \{$H_i$, $E_{\alpha}$, $E_{-\alpha}$ \} is the standard
   Cartan-Weyl basis of a semi-simple Lie algebra $g$ satisfying the standard
   commutation relations.

       Based on the Lie algebra $g$, we introduce a unitary operator
        
        \begin{equation}
         \label{1.1}
        W(\xi)=exp[\sum_{\alpha>0}(\xi_{\alpha}E_{\alpha}-\xi_{\alpha}^{*}E_{-\alpha})],
        \end{equation}
       in which $\xi_{\alpha}$ denote parameter functions to be determined.
        In fact, operator $W(\xi)$
        is a generalized displacement operator in the coset space
        $G/H$ [Zhang 1990]. Utilizing the
   Baker-Campbell-Hausdorff (B-C-H) formula
\begin{eqnarray}
e^{A}Be^{-A}=B+[A,B]+\frac{1}{2!}[A,[A,B]]+\cdot\cdot\cdot,
\nonumber
\end{eqnarray}
and taking advantages of the standard
        commutation relations of  semisimple Lie algebra $g$, the following
        identities can be obtained
         \begin{equation}
         \label{1.2}
         W(\xi)^{-1}G_{j}W(\xi)=\sum_{i}\gamma_{ij}(\xi)G_{i},
         \end{equation}
        where $G_{j}$ stand for the generators $H_i$ and $E_{\pm\alpha}$,
         $\xi$  denotes the set of $\{\xi_{\alpha}\}$,
         $\gamma_{ij}(\xi)$ are parameter functions.
         This is a central step in the overall treatment of the problem. Eq. (3)
         can directly be rewritten into the following form
        \begin{equation}
         \label{1.2}
         W(\xi)^{-1}HW(\xi)=\sum_i\eta_i(\xi,\varepsilon,\lambda)H_i+
         \sum_{\alpha>0}[(\mu_{\alpha}(\xi,\varepsilon,\lambda)E_{\alpha}
         +\mu^{*}_{\alpha}(\xi,\varepsilon,\lambda)E_{-\alpha}].
         \end{equation}
         The non-Cartan generators $E_{\pm \alpha}$ in the right-hand
          side are not diagonal and can be eliminated
       through setting
          \begin{equation}
           \mu_{\alpha}(\xi,\varepsilon,\lambda)=0
           \end{equation}
           for all parameters $\alpha$ which give constraints to the
         parameter-functions  $\{\xi_{\alpha}\}$ and  the parameters
         $\varepsilon,\lambda$. Then we get
         \begin{equation}
         HW(\xi)\mid Ref>=W(\xi)\sum_i\eta_i(\xi,\varepsilon,\lambda)H_i \mid Ref>.
         \end{equation}
        Here $\mid Ref >$ denote the reference states required to be the
      common eigenstates of all $H_i$.
      Therefore, $\sum_i\eta_i(\xi,\varepsilon,\lambda)H_i $
      can be substituted by functions acting on the reference states,
          i.e.,
           \begin{equation}
           HW(\xi)\mid Ref>=\omega(\xi,\varepsilon,\lambda,Ref) W(\xi)\mid Ref>.
           \end{equation}
        This is the eigen-equation of Hamiltonian
          Eq. (1), thus the algebraic diagonalization have been accomplished.
         State $W(\xi)\mid Ref>$ and $\omega$ are the energy eigenstates and
          eigenvalues of corresponding system, respectively.

          Now we briefly illuminate the completeness of the solution.
          For related fundamental introduction, please see reference [Zhang 1990, p877-879].
          In fact, the state $\mid Ref >$ and  the operator $W(\xi)$ here correspond to
          the state $\mid \Lambda,\Lambda >$ (the common eigenstate of $H_i$)
          and the displacement operator $\Omega$ in that reference, respectively.
          So, eigenstates $W(\xi)\mid Ref>$ correspond to states $\mid \Lambda, \Omega>$
           in that reference which play the role of complete basis of state space.

   \section {Squeezed number eigenstates of XYZ
   antiferromagnetics
   under an external magnetic field}
          The Hamiltonian of XYZ Heisenberg  antiferromagnetics
 under an external magnetic field  $\textbf{B}=B{\hat e}_z$ (along $z$ axis)
    is described by
      \begin{equation}
      H=-J\sum_{<i,j>}(\eta_{x} S_i^xS_j^x+\eta_{y}
      S_i^yS_j^y+
      S_i^zS_j^z)+\sum_{i}\textbf{B}\cdot\textbf{S}_i\;\;\;\;\;\;(J<0,\;\;\eta_{x},\eta_{y}>0).
       \end{equation}
    where the notation $<i,j>$ denote the nearest neighbor bonds.
     We employ the traditional two-sublattice treatment, i.e.,
     the spin directions are upwards for sites on sublattice A and downwards for
sublattice B, and then apply Holstein-Primakoff
       transformation [Holstein 1949]:
        $$
       S_a^z=-s+a^{\dag}a,\;\;\;\;\;\;\;\;s_b^z=s-b^{\dag}b,
        $$
        \begin{equation}
        S_a^{\dag}=(2s)^{\frac{1}{2}}a^{\dag}(1-a^{\dag}a/2s)^{\frac{1}{2}}
       ,\;\;\;\;
       S_b^{\dag}=(2s)^{\frac{1}{2}}(1-b^{\dag}b/2s)^{\frac{1}{2}}b,
         \end{equation}
         $$
         S_a^{-}=(S_a^{\dag})^{\dag},\;\;\;\;\;\;\;\;\;\;\;\;
          S_b^{-}=(S_b^{\dag})^{\dag},
          $$
    where $a^\dagger$, $a$ ($b^\dagger$, $b$) can be regarded as the creation and
    annihilation operators of boson on sublattice A (sublattice B).
    The particle numbers $a^{\dagger}a$, $b^{\dagger}b$ can't exceed $2s$.

    In low temperature and low excitation condition,
    $\langle a^{\dag}a \rangle,  \langle b^{\dag}b \rangle <<s$,
    so the non-linear interaction in Hamiltonian Eq. (8) can be reasonable ignored [Kittel
    1963]. Based on this  "big s" approximation,
    transforming the operators
     into momentum space, and leaving out the biquadratic terms,
     we get
           \begin{eqnarray}
        H &=& 2zsJ[Ns-\frac{1}{2} (\sum_{\bf k}{\cal H}_{\bf{k}}-1)],\\
       {\cal H}_{\bf{k}} &=& a _{\bf k}^{\dag}a_{\bf k}
       +a _{-\bf k}^{\dag}a_{-\bf k}+b_{\bf k}^{\dag}b_{\bf k}+b_{-\bf k}^{\dag}b_{-\bf k}+
       \upsilon_{\bf k}(a_{\bf k}b_{\bf
       {-k}}^{\dag}+a_{-\bf k}b_{\bf
       {k}}^{\dag}
       +  a_{\bf k}^{\dag}b_{\bf{-k}}+a_{-\bf
       k}^{\dag}b_{\bf{k}}) \nonumber\\
       & & +{\rho}_{\bf k}(a_{\bf k}b_{\bf k}+a^{\dag}_{\bf k}b^{\dag}_{\bf k}+a_{-\bf k}b_{-\bf k}+a^{\dag}_{-\bf k}b^{\dag}_{-\bf k})
       +\mu(b_{\bf k}^{\dag}b_{\bf{k}}-a_{\bf k}^{\dag}a_{\bf{k}}+b_{-\bf k}^{\dag}b_{-\bf{k}}-a_{-\bf k}^{\dag}a_{-\bf{k}}),
        \end{eqnarray}
        with
        \begin{equation}
        \upsilon_{\bf k}=\frac {\eta_{x}-\eta_{y}}{2}\gamma_{\bf k},\;\; {\rho}_{\bf k}= \frac
        {\eta_{x}+\eta_{y}}{2}\gamma_{\bf k},\;\;\gamma_{\bf k}=\frac{1}{z}\sum_{\bf R} e^{i{\bf k}\cdot
       \bf R},\;\;
        \mu=\frac{B}{2zsJ}.
        \end{equation}
       Here $\bf R$ is a vector connecting
        an atom with its nearest neighbor, and the sum
        runs over the $z$ nearest neighbors. $ \bf k$ is restricted in the Brillouin zone.
        2N is total number of the lattices.

        ${\cal H}_{\bf{k}}$ can be
        expressed as the linear combination of six generators of Lie algebra $so(3,2)$, i.e.,
         \begin{eqnarray}
        {\cal H}_{\bf{k}}=E_{3}^{\bf k}+\mu F_{3}^{\bf k}+{\rho}_{\bf k}(E_{+}^{\bf k}+E_{-}^{\bf k})+
        \upsilon_{\bf k}(F_{+}^ {\bf{k}}+F_{-}^{\bf k}),
        \end{eqnarray}
        where the  generators take the following forms
        \begin{eqnarray}
       \label{e5}
        & & E_{+}^{\bf k} =a^+_{\bf k}b^+_{\bf k}
        +a^+_{-\bf k}b^+_{-\bf k},
        \;\;\;\;\;\;\;\;\;\;\;\;\;\;\;\;\;\;\;\;\;\;\;\;
        F_{+}^{\bf k} =a_{\bf k}b^{+}_{-\bf k}
        +a_{-\bf k}b^+_{\bf k},
        \nonumber \\
        & & E_{-}^{\bf k} =a_{\bf k}b_{\bf k}
        +a_{-\bf k}b_{-\bf k}
        \;,\;\;\;\;\;\;\;\;\;\;\;\;\;\;\;\;\;\;\;\;\;\;\;\;
         F_{-}^{\bf k} =a^+_{\bf k}b_{-\bf k}
        +a^+_{-\bf k}b_{\bf k},
        \\
       & & E_{3}^{\bf k} =\frac{1}{2}(n^a_{\bf k}+n^a_{-\bf k}
        +n^b_{\bf k}+n^b_{-\bf k}+2),\,\,\;\;
         F_{3}^{\bf k} =\frac{1}{2} (n^b_{\bf k}+n^b_{-\bf k}
        -n^a_{\bf k}-n^a_{-\bf k}).
        \nonumber
          \end{eqnarray}
          The other four generators of $so(3,2)$ are
          \begin{eqnarray}
          U_{+}^{\bf k} = b^+_{-\bf k}b^+_{\bf k},
        \;\;\;\;\;
        U_{-}^{\bf k} = b_{\bf k}b_{-\bf k},
        \;\;\;\;\;
        V_{+}^{\bf k} = a^+_{-\bf k}a^+_{\bf k},
        \;\;\;\;\;
        V_{-}^{\bf k} = a_{\bf k}a_{-\bf k}.
        \end{eqnarray}
 The non-vanishing commutation relations of this algebra read
  \begin{eqnarray}
  \label{e6}
  & & [E_{+}, E_{-}]=-E_{3},
  \;\;\;\;
   [E_{3}, E_{\pm}]= \pm E_{\pm},
  \;\;\;\;
    [F_{+}, F_{-}]= F_{3},
     \;\;\;\;
     [F_{3}, F_{\pm}]= \pm F_{\pm},
  \nonumber \\
  & & [E_{3}, U_{\pm}]= \pm U_{\pm},\;\;\;\;[F_{3},U_{\pm}]= \mp U_{\pm},
  \;\;\;\;\;
    [E_{3}, V_{\pm}]= \pm V_{\pm},
  \;\;
    [F_{3}, V_{\pm}]= \pm V_{\pm},
  \nonumber \\
  & &  [E_{\pm}, V_{\mp}]= \mp F_{\mp},
  \;\;\;\;
   [F_{\pm}, U_{\pm}]= \pm E_{\pm} ,
  \;\;\;\;
    [E_{\pm}, F_{\pm}]= \mp V_{\pm}, \;[V_{+}, V_{-}]=-(E_{3} +
     F_{3}),
  \nonumber \\
  & & [F_{\pm}, V_{\mp}]= \mp E_{\mp},
  \;\;\;\;
    [E_{\pm}, U_{\mp}]= \mp F_{\pm}  ,
  \;\;\;\;
     [E_{\pm}, F_{\mp}]= \mp U_{\pm},\;[U_{+}, U_{-}]=-(E_{3} - F_{3}). \nonumber
  \end{eqnarray}
  One can see that \{$E_{+}^{\bf k},E_{-}^{\bf k},E_{z}^{\bf
   k}$\} forms an $so(2,1)\approx su(1,1)$ subalgebra, while \{$F_{+}^{\bf k},F_{-}^{\bf k},F_{z}^{\bf
   k}$\} forms an $so(3)\approx su(2)$ one. When $\eta_{x}=\eta_{y}$,
   the system will reduce to the $su(1,1)$ case [Xie 2002].

Hamiltonian Eq. (13) possesses the form of  Eq. (1), then  one can
do the diagonalization of Hamiltonian Eq. (13) through the
algebraic diagonalization method introduced in section II. Note
that the reference state $\mid Ref>$ is required to be the common
eigenstate of $E^k_{3}$ and $F^k_{3}$, $\mid Ref>$ should take the
form of number state $ \mid n^k_a,n^k_b>$. So, the eigenstates
$W(\xi)\mid Ref>$ are revealed to be squeezed number state, which
had been studied in quantum optics field [Yuen 1976; Kim 1989;
Nieto 1997].

In the calculation process of the diagonalization, we found that,
if a general full-parameter $W(\xi)$ is set, the infinite
progressions induced by the B-C-H formula can't replaced by
analytical functions. This lead to huge difficult even for number
calculation. For the purpose of studying the formal analytical
solutions and the general property of the squeezed number
eigenstate, we set
  \begin{equation}
   W(r,\theta)=\exp\{r[\cos\theta E_{+}
    +\sin\theta(V_{+}-U_{+})
   -h.c.]\}.
   \end{equation}
  Utilizing the commutation relations of $so(3,2)$ and the
   B-C-H formula,
  after a lengthy calculation,  we get
 \begin{eqnarray}
 W^{\dag}({r,\theta}){\cal H_{\bf k}} W({r,\theta}) =\omega^{\bf k}_{E}E_{3}+ \omega^{\bf k}_{F}F_{3},
 \end{eqnarray}
 where
 \begin{eqnarray}
 \omega_{E}^{\bf k}& = &
 \cosh2r+{\rho}_{\bf k} \cos\theta
 \sinh2r, \\
 \omega_{F}^{\bf k}& = &
 \mu(1+2\sin^2 \theta \sinh^2r)+ \upsilon_{\bf k}\sin 2\theta \sinh^2r,
 \end{eqnarray}
 with the constraint equations
 \begin{eqnarray}
  \label{e6}
   & &\cos \theta \sinh 2r+
{\rho}_{\bf k}(\cosh^2 r-\cos 2\theta \sinh^2r)=0,\\
  & & \mu \sin 2\theta\sinh^2 r+
 \upsilon_{\bf k}(\cosh^2 r+\cos^2 \theta\sinh^2 r)=0,\\
  & & \mu \sin \theta \sinh
  2r+{\rho}_{\bf k}\sin 2\theta \sinh^2 r
   -\upsilon_{\bf k}\cos \theta \sinh 2r =0.
   \end{eqnarray}
 The role of these equations is to eliminate the
 non-Cartan generators appearing originally on the right-side of Eq. (17).
 Note that there are 5 parameters in Eqs. (20-22), so
one of the parameters ${\rho}_{\bf k}, \upsilon_{\bf k}, \mu$
should be not independent. Acting both sides of Eq. (17) on
reference number state $|n^{\bf k}_a,n^{\bf k}_b\rangle $, we get
 \begin{eqnarray}
 {\cal H_{\bf k}} |R_{\bf k}>=(\omega_{a}^{\bf k} n^{\bf k}_{a}+\omega_{b}^{\bf k} n^{\bf
 k}_{b} +\frac{1}{2}\omega_{E}^{\bf k})|R_{\bf k}>,
 \end{eqnarray}
where
           \begin{equation}
           |R_{\bf k}>=W(r_{\bf k}, \theta_{\bf k})|n^{\bf k}_a,n^{\bf k}_b\rangle,
           \end{equation}
 \begin{eqnarray}
 \omega_{a}^{\bf k}=\frac{\omega_{E}^{\bf k}+\omega_{F}^{\bf k}}{2},\;\;\;
 \omega_{b}^{\bf k}=\frac{\omega_{E}^{\bf k}-\omega_{F}^{\bf k}}{2}.
 \end{eqnarray}

Eq. (23) is the eigen-equation.  The squeezed number state
$W(r_{\bf k}, \theta_{\bf k})|n^{\bf k}_a,n^{\bf k}_b\rangle$ is
the
  eigenstate of the system, and $\omega_{a}^k$ and
$\omega_{b}^k$ are the energies of the two different magnons
respectively. It is shown that the magnetic field lift the magnon
degeneracy. In the study of the number calculation, we find
usually their exist several solutions, but only one solution give
real energy eigenvalue. For instance, in a special case the
reasonable solution is, ${\rho}=0.1$, $\upsilon=-0.0006$,
$\mu=0.0383$, $\theta=1.2506$, $r=-0.1636$, $\omega_{a}=0.5418$,
$\omega_{b}=0.5017$.

Now we discuss the relation between the algebraic diagonalization
method used here and the traditional Bogoliubov-Valatin
transformation. To a certain extend, these two methods are
equivalent. The Bogoliubov-Valatin unitary transformation is known
to acts on single creation or annihilation operator. In our case,
the Lie algebra generators are composed of the quadratic form of
the creation and annihilation operators. So, the unitary
transformation acted on the Lie algebra
 generators corresponds to the product of two unitary
 transformations
 acting on single creation or annihilation
operator respectively. This fact connects the two methods, and
leads to that the eigenvalues should be the same and the energy
eigenstates can be connected through considerable transformation.
In fact, the unitary operator $W(r,\theta)$ (Eq. 16) provides a
two-mode squeezing transformation,
   \begin{eqnarray}
 \alpha_k &=& W^{-1}(r,\theta)a_k W(r,\theta)=a_k\cosh r+ (\cos \theta
b_k^\dag-\sin\theta a_k^\dag) \sinh r, \\
 \beta_k &=& W^{-1}(r,\theta)b_k W(r,\theta)=b_k\cosh r+(\cos \theta
a_k^\dag+\sin\theta b_k^\dag) \sinh r.
\end{eqnarray}
Here $\alpha_k$ and $ \beta_k$ are new bosonic operators. In fact,
the anti-transformation of Eq. (26,27) can just be regarded as a
corresponding Bogoliubov transformation. For briefness, we would
not write them out here. The squeezed number form is
$(\omega_{a}^{\bf k} n^{\bf k}_{a}+\omega_{b}^{\bf k} n^{\bf
 k}_{b} +\frac{1}{2}\omega_{E}^{\bf k})W(r_{\bf k}, \theta_{\bf k})|n^{\bf k}_a,n^{\bf
 k}_b\rangle$, where the $n^{\bf k}_a,n^{\bf
 k}_b$ are the eigenvalues of the particle number operators of  $a^\dag_k a_k, b^\dag_k b_k$.
 The corresponding Bogoliubov form is
$[{\omega^{\bf
k}_\alpha}(\alpha^\dag_k\alpha_k+\frac{1}{2})+{\omega^{\bf
k}_\beta}(\beta^\dag_k \beta_k+\frac{1}{2})]|0,0\rangle$, where
the $\alpha_k$ and $ \beta_k$ is the bosonic operators of the
quasi-particles. It's apparent that the two
 methods work in different operator space, one is for "quasi-particle", the other is for "real
 particle".

\section {Some statistical properties of eigenstate}
 In this section, we discuss
some statistical properties of the eigenstate as squeezed number
state. The average magnetization of sublattice A and B take the
forms
  \begin{eqnarray}
  <s_B^z>&=&Ns-\sum_{\bf k} <b^{\dag}b>_{\bf k},\\
  <s_A^z>&=&-Ns+\sum_{\bf k} <a^{\dag}a>_{\bf k},
  \end{eqnarray}
  where,
    \begin{eqnarray}
  <b^{\dag}b>_{\bf k}=\frac{1}{2}\cosh2r(n_a^{\bf k}+n_a^{-{\bf k}}+n_b^{\bf k}+n_b^{-{\bf k}}+2)+
  \frac{1}{2}(1+2\sin^2\theta\sinh^2r)(n_b^{\bf k}+n_b^{-{\bf k}}-n_a^{\bf k}-n_a^{-{\bf k}})-1,    \nonumber
  \end{eqnarray}
   \begin{eqnarray}
  <a^{\dag}a>_{\bf k}=\frac{1}{2}\cosh2r(n_a^{\bf k}+n_a^{-{\bf k}}+n_b^{\bf k}+n_b^{-{\bf k}}+2)+
  \frac{1}{2}(1+2\sin^2\theta\sinh^2r)(n_a^{\bf k}+n_a^{-{\bf k}}-n_b^{\bf k}-n_b^{-{\bf k}})-1.    \nonumber
  \end{eqnarray}
When the particle numbers $n^{\bf k}_a$ and $n^{\bf k}_b$ vanish,
the eigenstate becomes squeezed vacuum state. In this state,
\begin{eqnarray}
  <b^{\dag}b>_{\bf k}= <a^{\dag}a>_{\bf k}=\cosh2r-1.
  \end{eqnarray}
   One can see, even in the squeezed vacuum state, i.e., the ground state, the spin
   direction of different sites on a sub-lattice is not necessarily  the same; this
property of disordering has relations with the zero-point motion
of the magnon in fact. Moreover, in the squeezed vacuum state, the
influence of the external magnetic field upon the agree of average
spin reversal of the two sub-lattice is inviolable the same; while
in other squeezed number states, this property becomes true only
in the condition that $n_a^{\bf k}+n_a^{-{\bf k}}=n_b^{\bf
k}+n_b^{-{\bf k}}$.

  Now we study the other quantum
 effects from the point view of quantum optics.
For simplicity, hereafter only squeezed vacuum state is mentioned.
The second-order correlation
 functions  of the two excitation modes can be readily obtained as
    \begin{eqnarray}
   {g^{{\bf k}}_{b}}^{(2)}&=&{g^{{\bf k}}_{a}}^{(2)}= \frac {\langle {a^{\dag}_{{\bf k}}}^2 {a_{{\bf k}}}^2 \rangle}
                      {{\langle a^{\dag}_{{\bf k}} a_{{\bf k}}
                      \rangle}^2}=2+\sin \theta\coth^2r,  \\
    {g^{{\bf k}}_{ab}}^{(2)}&=&\frac {\langle a^{\dag}_{\bf k}a_{\bf k}b^{\dag}_{\bf k}b_{\bf k} \rangle}
                      {\langle a^{\dag}_{{\bf k}}a_{{\bf k}} \rangle \langle b^{\dag}_{{\bf k}}b_{{\bf k}}
                      \rangle}=1+2\cos\theta \coth^2r.
  \end{eqnarray}

  The Mandel $Q$ parameters [Mandel 1979;1986] read
  \begin{eqnarray}
   Q^{\bf k}_{b}= Q^{\bf k}_{a}= \frac {\langle (\Delta a^{\dag}_{\bf k}a_{\bf k})^2  \rangle}
                      {{\langle a^{\dag}_{\bf k}a_{\bf k} \rangle}}-1 =\sinh^2 r+ \sin \theta \cosh^2r.
                       \end{eqnarray}
Eq. (33) illustrates that in squeezed vacuum state, each of the
three statistics,
  sub-Poissonian ($Q> 0$), Poissonian ($Q=0$), super-Poissonian ($-1 \leq Q<0$,
  nonclassical state)
  are all possible to exist in the two magnon
 modes, relying on the values of the parameters
 $r$ and $\theta$, which are determined by the parameters
 $\upsilon_{\bf k}$, ${\rho}_{\bf k}$, $\mu$.

If the parameter $I^{\bf k}=\sqrt{{g^{\bf k}_b}^{(2)}{g^{\bf
k}_{a}}^{(2)}}/{g^{\bf k}_{ab}}^{(2)}-1<0$, then the
Cauchy-Schwartz
 inequality (CSI) [Agarwal 1988] is violated,
 and the correlation between the two {\bf k}-magnon modes
 is nonclassical. In squeezed vacuum state,
 \begin{eqnarray}
  I^{\bf k}=\frac{1+(\sin\theta-2\cos\theta)\coth^2r}{1+2\cos\theta\coth^2r},
   \end{eqnarray}
 it can be found that larger squeezing
 parameter $r$ leads to smaller possibility  of the violation of CIS,
 i.e. the achievement of nonclassical
 correlation. This is not surprising since larger squeeze
 parameter $r$
 corresponds to larger particle number, tending to approaching classical case.

  \section {connection with the two-dimension coupled harmonic oscillators}
 For two different physical systems, if their Hamiltonian
 can be written into the same combination of the generators of a Lie
 algebra completely through different realizations of this Lie algebra,
 it's reasonable to believe that the two systems possess the same
 eigenvalue structure. Now we consider the relation of
  XYZ antiferromagnetic Heisenberg model under an
external magnetic field and two-dimension coupled harmonic
oscillators.
 Since the $so(3,2)$ algebra can also be constructed  in the form
 \begin{eqnarray}
  \label{e4}
  & & E_{+} = a^{\dag}_{1}a^{\dag}_{2}
 ,\;\;
  E_{-} = a_{1}a_{2}
  ,\;\;
   E_{3} =\frac{1}{2}(a^{\dag}_{1}a_{1}+a^{\dag}_{2}a_{2}+1),
  \nonumber \\
  & &  F_{+} = a_{1}a^{\dag}_{2}
  ,\;\;
   F_{-} = a^{\dag}_{1}a_{2}
  ,\;\;
   F_{3} =\frac{1}{2} (a^{\dag}_{2}a_{2}-a^{\dag}_{1}a_{1}),\\
   & & U_{+} =\frac{1}{2} {a^{\dag}_{1}}^2
  ,\;\;
   U_{-}=\frac{1}{2} {a_{1}}^2,\;\;
   V_{+}=\frac{1}{2} {a^{\dag}_{2}}^2
  ,\;\;
   V_{-}=\frac{1}{2} {a_{2}}^2. \nonumber
  \end{eqnarray}
Eq. (13) can also be realized as
  \begin{equation}
 {\cal H}'=\frac{1-\mu}{2}a_1^{\dag}a_1+\frac{1+\mu}{2}a_2^{\dag}a_2+\rho(a_1
a_2+ a_1^\dag a_2^\dag)+\upsilon (a_1 a_2^\dag+a_1^\dag a_2).
 \end{equation}
 Hereafter we leave out the index ${\bf k}$ for all the parameters
and operators for convenience. So, for every ${\bf k}$ modes,
   utilizing the realization $
 a_j=(m\omega_j x_j+i p_j)/\sqrt{2m\omega_j},
  $
 we can map a Hamiltonian of two-mode coupled harmonic oscillators  corresponding hamiltonian ${\cal H}$ of Eq. (11),
 \begin{equation}
 {\cal H}'=\sum_{j=1,2}[\frac{p_{j}^2}{2m}+\frac{\omega_{j}^2}{2} m
 x_j^2]+\lambda_1\omega_1\omega_2 m x_1x_2+\lambda_2\frac{p_1p_2}{m}
 ,
 \end{equation}
 with
 \begin{eqnarray}
    \frac{\sqrt{\omega_1
    \omega_2}}{2}(\lambda_1-\lambda_2)=\rho,\;\;\;
  \frac{\sqrt{\omega_1
  \omega_2}}{2}(\lambda_1+\lambda_2)=\upsilon,\;\;\;
 1-2\omega_1=2\omega_2-1=\mu.
 \end{eqnarray}

 Hamiltonian ${\cal H}$ and ${\cal H}'$ should have the same energy
 spectrum, i.e., in low excitation case, every $\bf k$ modes for XYZ
antiferromagnetics in external magnetic field can be mapped onto a
two-dimension coupled harmonic oscillators with $x$-$x$ and
$p$-$p$ coupling, with the corresponding relation of the
parameters
 Eq. (38). One can see that, the system parameters $\upsilon$ and
 $\rho$ are linked  to the coefficients of $x$-$x$ and
$p$-$p$ coupling; while the role of the parameters $\mu$ (external
magnetic field term) is to lift the energy degeneracy of the
corresponding harmonic oscillators.

\section {Concluding remarks}

In this paper, based on the introduction of an algebraic
diagonalization method, the eigenstates of XYZ antiferromagnetics
are revealed to be the number squeezed states in the framework of
linear spin-wave approximation. Different from the traditional
Bogoliubov-valatin transformation, we study the problem using the
algebraic diagonalization method. The characteristic of this
method lies in that, firstly, one can do calculation in virtue of
the commutation relations of Lie algebra instead of the Heisenberg
algebra; Secondly, the energy eigenstates are revealed to be
squeezed number states, which had been studied in quantum optics
field [Yuen 1976; Kim 1989; Nieto 1997], making it convenient to
study the physical properties of the system. It is shown that, the
presence of magnetic field results in the non-degeneracy of the
two excited magnon; even in the squeezed vacuum state, the spin
direction on a sub-lattice is not necessarily the same (this
disordering in a certain extent is attributed to the zero-point
motion of the spin oscillator). Moreover, in the squeezed vacuum
state, the influence of the external magnetic field upon the agree
of average spin reversal of the two sub-lattice is absolutely the
same. Some statistical property of the squeezed vacuum state, the
second-order correlation
 functions,  Mandel $Q$ parameters, violation of CIS are
 also discussed. By virtue of the algebra method, we illuminate that,
 XYZ antiferromagnetics under an external magnetic field
 can be mapped onto a two-dimension coupled harmonic oscillators with
$x$-$x$ and $p$-$p$ coupling. Thus, the properties of the
two-dimension coupled harmonic oscillators with $x$-$x$ and
$p$-$p$ coupling, can readily be connected to the XYZ
antiferromagnetics under an external magnetic field through
adjusting parameter. Besides, our method can also be used to treat
the sub-ferromagnetic case, and the case of existing anisotropic
crystal magnetic field. The application of our results in related
physical fields, as well as  the realization
and employment of squeezed number states in other physical systems, leave for further studies.\\

\noindent ACKNOWLEDGMENT. This work is supported by the National
Science Foundation of China (11447103), Education Department of
Beijing Province and Beihang University.

\end{document}